\documentclass[%
 reprint,
 amsmath,amssymb,
 aps,
floatfix,
]{revtex4-2}

\usepackage{comment}
\usepackage{graphicx}% Include figure files
\usepackage{dcolumn}% Align table columns on decimal point
\usepackage{bm}% bold math
\usepackage{float}
\usepackage{color}
\usepackage{hyperref}% add hypertext capabilities
\hypersetup{
    colorlinks=true,
    linkcolor=blue,
    citecolor=blue,
    urlcolor=blue,
}
\renewcommand{\exp}[1]{\text{e}^{#1}}
\renewcommand{\vec}[1]{\boldsymbol{#1}}

\begin{document}

\preprint{APS/123-QED}

%\title{Multiscale approach for relaxation and characterization of moire systems}% Force line breaks with \\
\title{Accurate force-field methodology capturing atomic reconstructions in transition metal dichalcogenide moiré systems}% Force line breaks with \\

\author{Carl Emil Mørch Nielsen}
\author{Miguel da Cruz}
\author{Abderrezak Torche}
\author{Gabriel Bester}
\affiliation{%
 Institute of Physical Chemistry, University of Hamburg, 22607 Hamburg, Germany
}%

\date{\today}% It is always \today, today,
             %  but any date may be explicitly specified

\begin{abstract}

In this work, a generalized force-field methodology for the relaxation of large moiré heterostructures is proposed. The force-field parameters are optimized to accurately reproduce the structural degrees of freedom of some computationally manageable cells relaxed using density functional theory. The parameters can then be used to handle large moiré systems. We specialize to the case of 2H-phased twisted transition-metal dichalcogenide homo- and heterobilayers using a combination of the Stillinger-Weber intralayer- and the Kolmogorov-Crespi interlayer-potential. Force-field parameters are developed for all combinations of MX$_2$ for $\text{M}\in\{\text{Mo},\text{W}\}$ and $\text{X}\in\{\text{S},\text{Se},\text{Te}\}$. The results show agreement within 20 meV in terms of band structure between density functional theory and force-field relaxation. Using the relaxed structures, a simplified and systematic scheme for the extraction of the interlayer moiré potential is presented for both R- and H-stacked systems. We show that in-plane and out-of-plane relaxation effects on the moiré potential, which is made both deeper and wider after relaxation, are essential. An interpolation based methodology for the calculation of the interlayer binding energy is also proposed. Finally, we show that atomic reconstruction, which is captured by the force-field method, becomes especially prominent for angles below 4-5$^\circ$, when there is no mismatch in lattice constant between layers.
\end{abstract}

%\keywords{Suggested keywords}%Use showkeys class option if keyword
                              %display desired
\maketitle

%\tableofcontents

\section{\label{sec:introduction}Introduction}

\begin{figure}[ht]
\centering
\includegraphics[width=0.95\linewidth]{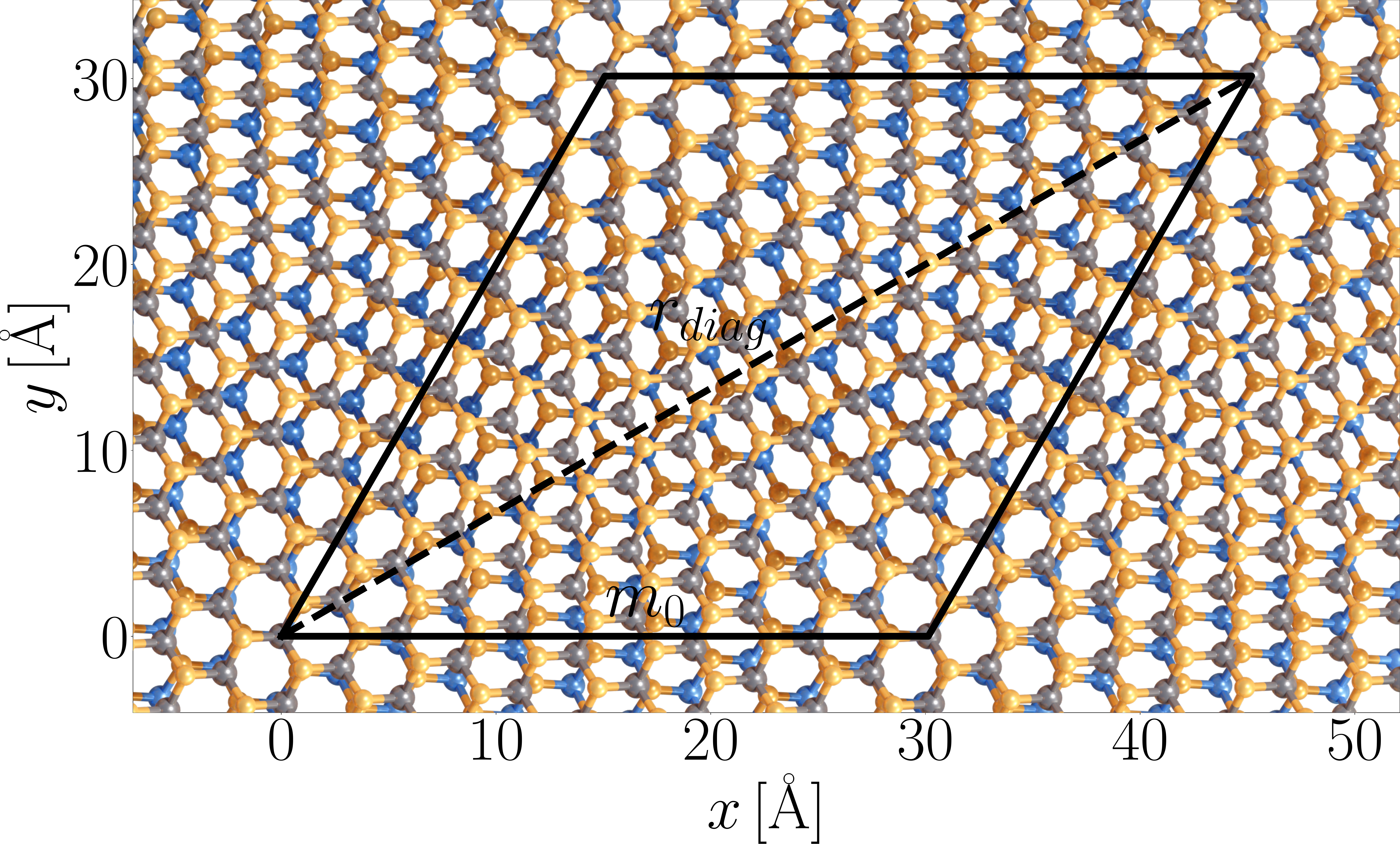}
\caption{\label{fig:moire_whole} WS$_2$ on MoS$_2$ twisted at an angle of $6.0^{\circ}$. Sulphur atoms are shown in orange, molybdenum in blue, and tungsten in grey. The moiré unit cell is shown with black solid lines, and has a moiré period, $m_0$, of 30.1 Å encompassing 546 atoms. The long diagonal, $r_{diag}$, is marked with a dashed black line.}
\end{figure}

Two dimensional (2D) moiré systems are currently an especially attractive playground for new technological applications \cite{Liu2016,Novoselov2016,Gibertini2019,Yao2021}. Lattice mismatch combined with the twist angle between the constituent layers allows for an ingenious way of external mechanical control of the moiré period and thus the resulting electronic properties. Without a doubt, the pioneering discovery of twisted bilayer graphene and its magic angle of 1.05$^{\circ}$ \cite{Bistritzer2011} was the major driving force toward the study of 2D heterostructures and constituted the basis for the field of twistronics. An interesting and widely studied class of moiré systems is the 2D family of transition metal dichalcogenides (TMDs), featuring strong light-matter interaction and large spin-orbit coupling with a sizable bandgap \cite{Al-Ani2022}. A fundamental advantage of TMDs is that flat minibands are not only realised at specific angles, but exist in a continuum of small angles \cite{Naik2018}. An example of a moiré-structured TMD-system can be seen in Fig. \ref{fig:moire_whole}. Moreover, experimental and theoretical findings of the excited states in type-II aligned heterostructured TMDs show evidence of spatially indirect excitons localized within certain registries of the moiré structure \cite{Choi2020,Baek2020,Naik2018,Guo2020}.  Moiré structured TMDs provide a platform for studying correlated quantum phenomena \cite{WuTutuc2018} including hole Mott insulator states at integer and fractional fillings with generalized Wigner crystallization, essentially creating a Fermi-Hubbard system \cite{Tang2020,Pan2020_1,Slagle2020,Pan2020_2,Regan2020}. Moiré structured TMD bilayer systems also allows for realization of Bose-Hubbard physics with excitons trapped in a periodic triangular potential and subject to strong Coulomb interactions \cite{Gotting2022}. 

Moiré physics in TMDs are largely determined by the shape of the twist-induced moiré potential, which arises from local stacking configurations, lattice corrugation and, for small angles, atomic reconstruction \cite{Rosenberger2020,Carr2018,Maity2021,Naik2018,Enaldiev2020,LiLi2021,Weston2020,Shabani2021,Sung2020}. As a consequence, relaxation effects are important for numerical simulations that involve moiré structured TMD systems prone to atomic reconstruction, and/or structures with a moiré period large enough to corrugate the individual layers \cite{Liu2021,Geng2020}. From an \textit{ab initio} standpoint, this presents a large challenge owing to the fact that relaxation is a computational bottleneck in such calculations. In an excellent paper by Naik \textit{et al.} \cite{Naik2019}, a method to overcome this problem is suggested by using a force-field model based on a combination of the Stillinger-Weber (SW) \cite{Stillinger1985,Zhou2017} and Kolmogorov-Crespi (KC) \cite{Kolmogorov2005,Kolmogorov2000} potentials. The SW force-field accurately describes the intralayer forces, while the KC potential captures van der Waals (vdW) interaction between layers and includes a stacking-dependent term. Previously, this had been parametrized and applied to graphene and hexagonal boron nitride \cite{Maaravi2017,Leven2016,Leven2014,Ouyang2018}, but is now also available for MX$_2$ homobilayers, where $\text{M}\in\{\text{Mo},\text{W}\}$ and $\text{X}\in\{\text{S},\text{Se}\}$ \cite{Naik2019,Ouyang2021}.

However, the parameters presented in Ref. \cite{Naik2019} are somewhat inaccurate when comparing to density functional theory (DFT) calculated results, e.g. for some structures, the bandgap is inaccurate by up to 100 meV. Even more importantly, the band curvature and energetic position of e.g. the lowest conduction band and highest valence band are skewed on similar scales. In Ref. \cite{Naik2019}, the parameters are developed by fitting to DFT binding energies which will not guaranty the force-field model to reproduce the DFT relaxed structure. In this work, the structural parameters of the DFT optimized structures (i.e. atomic positions and unit cell size) are used directly as target values for the optimization of the force-field parameters. Furthermore, the KC-parametrization of Ref. \cite{Naik2019} is presented on a \textit{per interaction basis}, meaning that atom-atom interactions are considered the same for different systems, e.g. S-S parameters for MoS$_2$- and WS$_2$-bilayers are the same. However, from a fundamental point of view, vdW interaction, being of long-range nature, is known to be sensitive to the surrounding environment. As such, we reparametrize the KC-potential on a \textit{per system basis}, which yields more accurate band structures. Furthermore, we expand the set of parameters to include heterobilayers with and without lattice mismatch, essentially covering all bilayer combinations of 2H-phased MX$_2$ for $\text{M}\in\{\text{Mo},\text{W}\}$ and $\text{X}\in\{\text{S},\text{Se},\text{Te}\}$. However, the method presented here is, in principle, extendable to any 2D moiré structure and not limited to TMDs. Our force-field parameters, along with a variety of relaxed structures can be found via Ref. \cite{ff_params}.

Lastly, we present two interpolation-based schemes to describe the \textit{interlayer exciton moiré potential} of lattice-matched heterostrutures with type-II band alignment by using a combination of the force-field method and DFT, which provides easy access to the potential for almost any angle. We extend this analogy to the binding energy, which allows for visualization of atomic reconstruction and the rate at which the reconstructed domains form with decreasing twist angle. Specifically, we see that atomic reconstruction becomes significant for angles below 4-5$^\circ$ for the TMD heterostructures studied here.

\section{\label{sec:Methodology}Methodology}

The first step is to develop the SW-parameters, which is done by considering the constituting monolayers one at a time. For 2H-phased TMD monolayers, the hexagonal symmetry reduces the structural degrees of freedom into two (target) parameters only, namely the lattice constant, $a_0$, and the \textit{intralayer} distance, $d_{intra}$, i.e. the out-of-plane X-X distance. Therefore, the SW-parametrization is carried out using $a_0$ and $d_{intra}$ as targets and reproduces them extremely well. The force-field relaxations are performed using the LAMMPS package \cite{Thompson2022}, and the optimization of parameters is carried out with use of the Dakota package \cite{Adams2009}.

For the optimization of the KC parameters we are following two strategies, depending on whether the constituting layers are lattice matched or not. 

\subsection{Lattice matched bilayers}

\begin{figure}[ht]
\centering
\includegraphics[width=\linewidth]{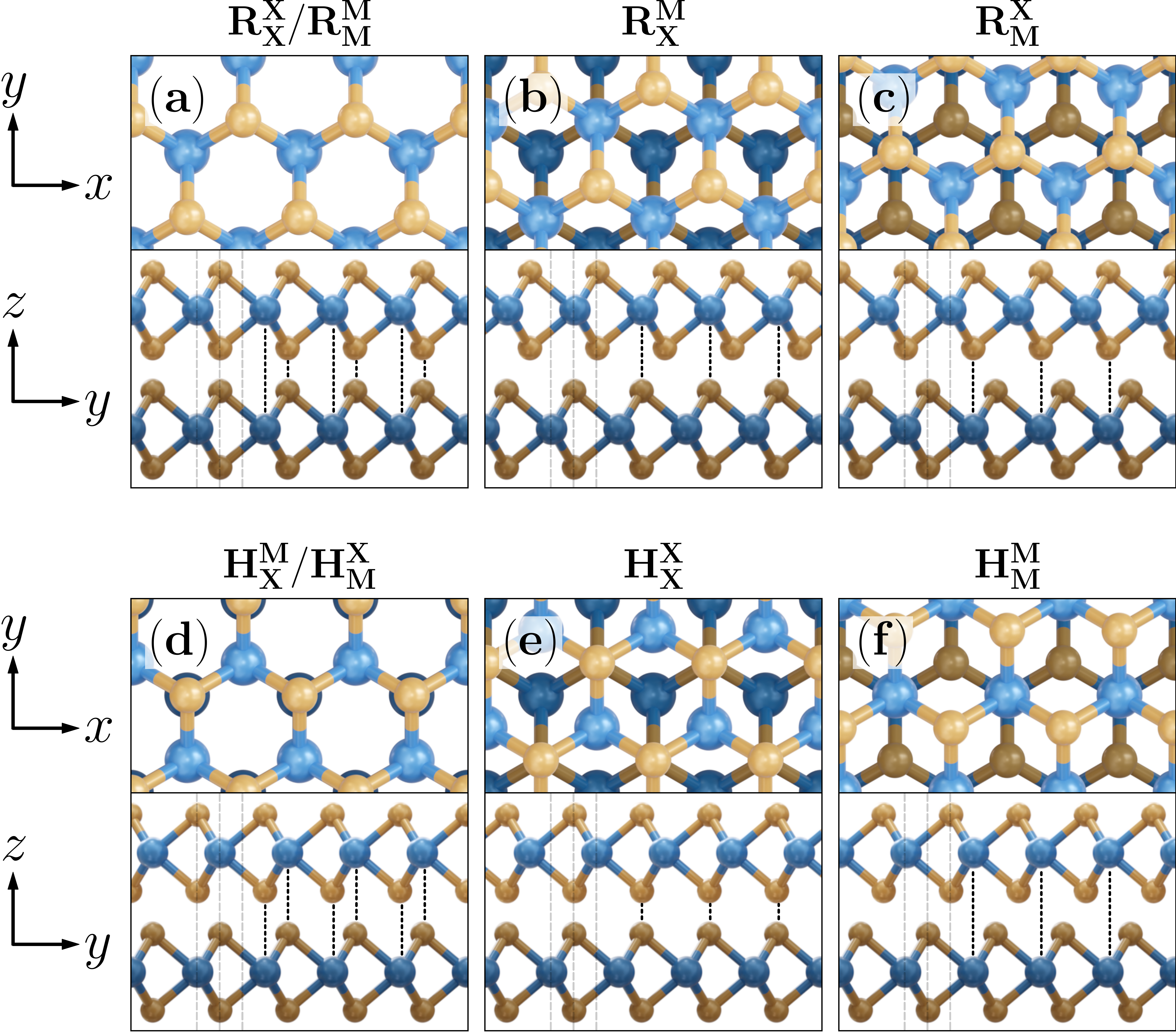}
%\label{fig:stackings}
\caption{\label{fig:stackings}The six high-symmetry stacking configurations of a bilayer with no lattice-mismatch. (a)-(c) ((d)-(f)) belong to the R(H)-stacking group. The stacking $\mathrm{R_X^X}/\mathrm{R_M^M}$ is also referred to as AA-stacking, and $\mathrm{H_X^M}/\mathrm{H_M^X}$ as AB-stacking. The dotted black lines indicate atoms that coincide along $z$, justifying the naming convention.}
\end{figure}

Bilayers that have the same chalcogen atom have a lattice constant mismatch $\delta\sim0.1\%$ and are treated as lattice matched. In this case, only one additional structural parameter is considered, namely the \textit{interlayer} spacing, $d_{inter}$ (M-M distance).
The KC-parameters are obtained by fitting to $a_0$, $d_{intra}$ and $d_{inter}$ for the six high-symmetry stacking configurations (HSSCs), while keeping the SW-parameters fixed. 
The HSSCs are depicted in Fig. \ref{fig:stackings} and are divided in two groups, namely R- and H-stacking, which differ by a rotation of one of the layers by 60$^\circ$.
This procedure follows the idea that the mechanical properties of the single layer is well described by the SW potential and is not altered by the interlayer interaction (KC potential). It is crucial to derive a force-field that is transferable between the different stackings since the twisted bilayers correspond to combinations of three different stackings, as will be demonstrated subsequently.

Moreover, as we will indirectly show in Sec. \ref{sec:HSIM}, every subcell of a lattice-matched moiré unit-cell is, to a certain extent, well described by a superposition of the HSSCs. Note, that this is not the case for lattice-mismatched systems where no local HSSCs can be identified. Justification of our methodology becomes trivial for smaller angles, where domains of the HSSCs make up a large fraction of the moiré unit cell. Finally, the small unit cells constructed with merely six atoms, makes both DFT calculations and the optimization schemes of Dakota and LAMMPS relatively fast. 

\subsection{Lattice mismatched bilayers}

In Table \ref{tab:mismatch} we show the lattice mismatch $\delta$ for the different combinations of chalcogen atoms (the metal atom is nearly irrelevant for the lattice constant). 
The lattice-mismatch of the systems investigated here (X=S,Se,Te) is so large that the construction of small six atom unit cells as done in the lattice matched case is not meaningful. The in-plane strain will radically change the electronic properties \cite{Shi2013}.
\begin{table}[hb]
\caption{\label{tab:mismatch} Angles chosen for fitting lattice-mismatched structures accompanied by the lattice-mismatch ($\delta$, found using DFT), number of atoms ($n_{atom}$), and the moiré lattice constant $m_0$ (moiré period).}
\begin{ruledtabular}
\begin{tabular}{cccccc}
X$^1$ & X$^2$ & $\theta$ ($^\circ$) & $\delta$ (\%) & $n_{atom}$ & $m_0$ (nm) \\
\hline
S  & Se & $5.68$ & $ 4.1 \pm 0.1$ & 525 & 3.0 \\
Se & Te & $5.07$ & $ 7.0 \pm 0.1$ & 471 & 3.0 \\
S  & Te & $0.00$ & $11.3 \pm 0.1$ & 543 & 3.2 \\
\end{tabular}
\end{ruledtabular}
\end{table}

\begin{figure}[ht]
\centering
\includegraphics[width=\linewidth]{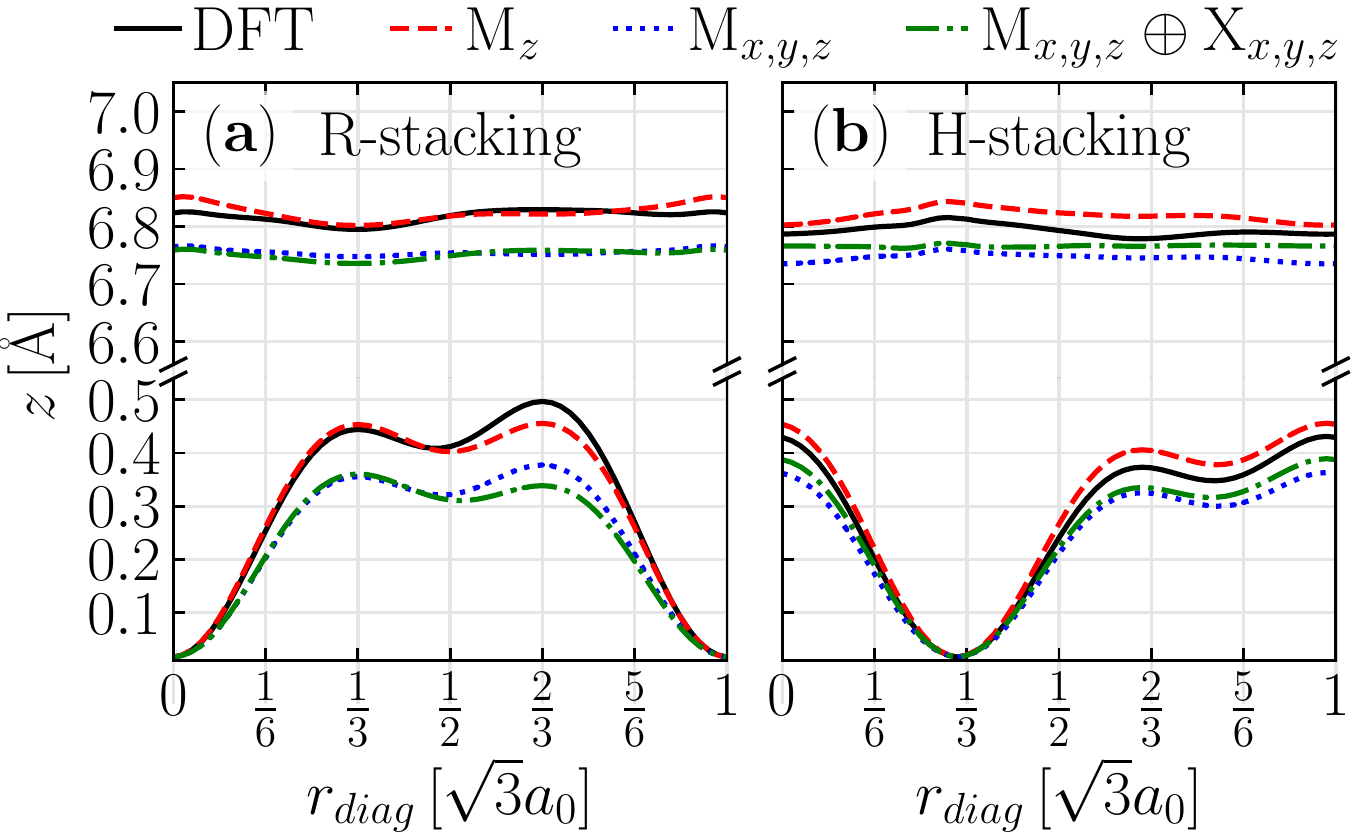}
%\label{fig:stackings}
\caption{\label{fig:lmmfit} $z$-coordinates of the metal atom in a 5.68$^{\circ}$ twisted WS$_2$-MoSe$_2$ bilayer along the long diagonal of the moiré unit cell, $r_{diag}$. The bottom curvy layer corresponds to WS$_2$, and the top more rigid layer corresponds to MoSe$_2$. The solid black curve is the DFT relaxed structure. The dashed red, dotted blue, and dashdotted green curves are force-field relaxed with KC-parameters (see text). (a) and (b) show R- and H-stackings, respectively.}
\end{figure}

To circumvent this problem, we use relatively small (about 500 atoms) moiré structures as targets for lattice-mismatched systems (See Table \ref{tab:mismatch}). Due to the reduced symmetry of lattice-mismatched systems, the only valid targets are the coordinates of all atoms of the moiré unit cell combined with the lattice constant. However, using all atomic coordinates, i.e. three spatial dimensions for each atom, renders the mesh adaptive search scheme for optimizing the KC-parameters infeasible, as the number of target values greatly exceeds the number of fitting parameters (Fig.~\ref{fig:lmmfit}, dashdotted green curve). As such, it is necessary to reduce the number of target values. However, considering only the three spatial coordinates of the metal atoms, thus reducing the target space by one third, also yields sub-optimal KC-parameters (Fig.~\ref{fig:lmmfit}, dotted blue curve). Lastly, optimizing only for the $z$-coordinates of the metal atoms, which further reduces the target space by one third, results in a much better fit (Fig. \ref{fig:lmmfit}, dashed red curve). As such, we ultimately choose the $z$-coordinates of the metal atoms and the lattice constant as target values for lattice-mismatched systems, which yields satisfactory KC-parameters, as discussed in Sec. \ref{sec:results}.

\subsection{The Kolmogorov-Crespi Potential}

As mentioned previously, the KC potential, $V_{ij}$, is intended to model interlayer effects between atom $i$ in one layer and atom $j$ in another, and is given by
\begin{align}
    V_{ij} &= \exp{-\lambda(r_{ij}-z_0)}\left[C+f(\rho_{ij})+f(\rho_{ji})\right]-A\left(\frac{r_{ij}}{z_0}\right)^{-6}, \nonumber
    \\
    \rho_{ij}^2 &= r_{ij}^2-(\vec{n}_i\vec{r}_{ij})^2, \nonumber
    \\
    \rho_{ji}^2 &= r_{ij}^2-(\vec{n}_j\vec{r}_{ij})^2, \nonumber
    \\
    f(\rho) &= \exp{-(\rho/\delta)^2}\sum_{n=0}^{2}C_{2n}(\rho/\delta)^{2n}.
\end{align}
$\vec{n}_i$ and $\vec{n}_j$ are the surface normals of the atom site $i$ and $j$ in each layer. The choice of neighbors used to determine the surface normals are the six nearest atoms in the respective strata (sublayer of the monolayer). The last term of $V_{ij}$ contains the $r^{-6}$ vdW dependence, and the first term has an exponentially decaying repulsion reflecting interlayer wave-function overlap. The square bracket functions contain a stacking dependent term, in contrast to e.g. the Lennard-Jones potential \cite{Kolmogorov2005}. As seen, $V_{ij}$ leaves in total eight parameters to be fitted. As mentioned in Ref. \cite{Naik2019}, it is possible to approximate $\vec{n}_{i,j}=\hat{z}$ corresponding to completely rigid layers, however, we do not make use of this approximation in order to capture more accurately the corrugation caused by the relaxation.

\subsection{Computational details}

We parametrize the potentials with different combinations of exchange correlation plus vdW correction. We find that using PBE \cite{Perdew1996} from PseudoDojo \cite{vanSetten201839,Hamann2013} with Grimme's DFT-D3 vdW correction \cite{Grimme2010} plus Becke-Johnson damping \cite{Becke2006} is best suited for parametrization. The structures are relaxed with QuantumEspresso \cite{Giannozzi2009,Giannozzi2017} using a $k$-space density of $15\times15$ ($1\times1$) for high-symmetry (moiré) unit cells. DFT computations of moiré systems are performed without spin-orbit coupling (SOC) to save computational resources, since they are only used for comparing DFT to SW+KC relaxed structures. We find that the lattice constant only converges at a cut-off energy of 40 Ha in all cases. More importantly, the chosen cut-off energy should be consistent between monolayers, untwisted bilayers and moiré structured bilayers, when comparing DFT to SW+KC. We use the modified SW implementation in LAMMPS for ease of use. For optimization in Dakota, we apply a mesh adaptive direct search algorithm starting from the parameters presented in Ref. \cite{Naik2019}. 

\section{\label{sec:results}Results}

\begin{figure}[ht]
\centering
\includegraphics[width=0.99\linewidth]{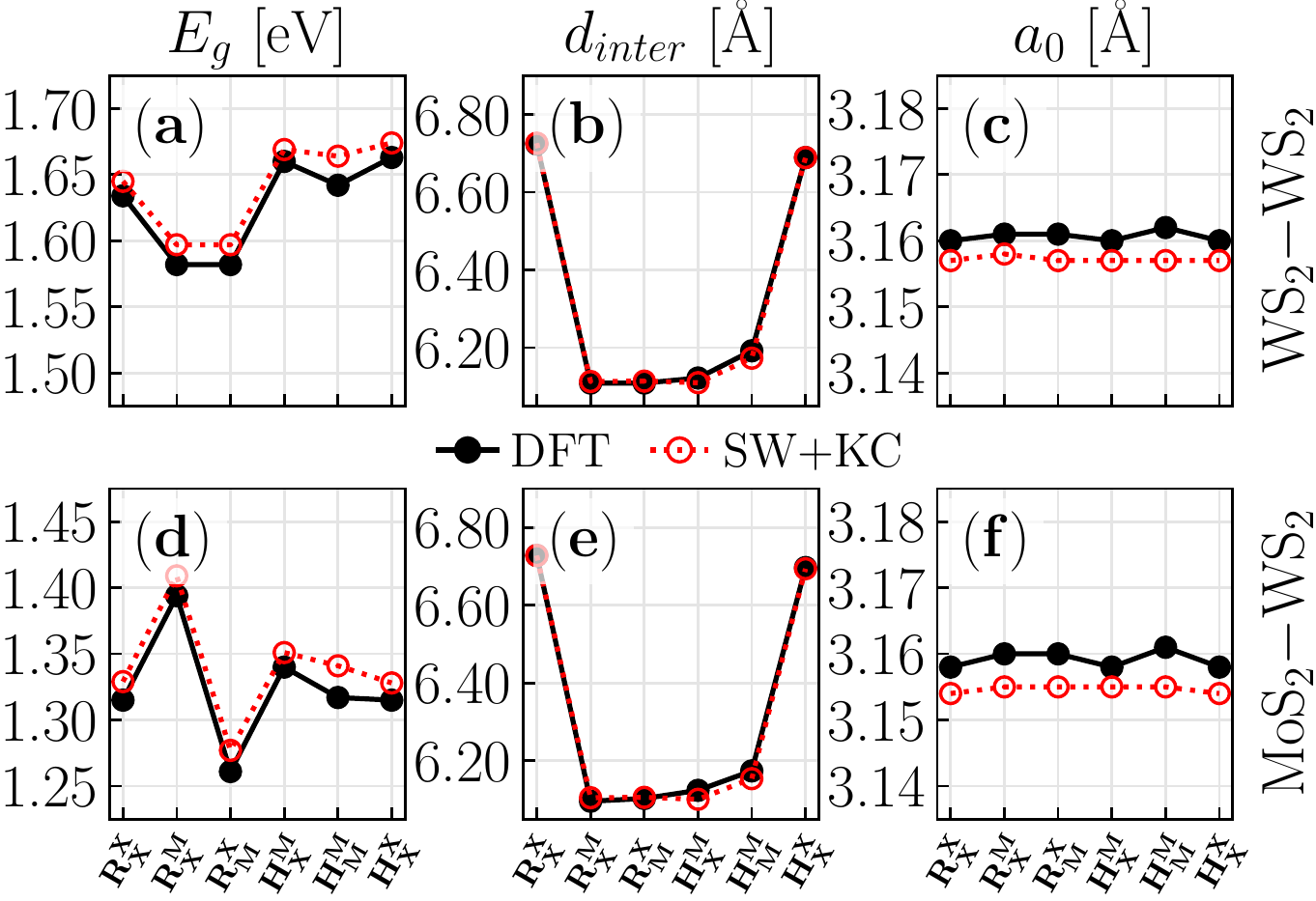}
\caption{\label{fig:lammps_fit}Comparison of $E_g$ at $\vec{K}_{\pm}$ in (a) and (d), interlayer spacing ($d_{inter}$) in (b) and (e), and lattice constant $a_0$ in (c) and (f) for the six HSSCs of a WS$_2$ homobilayer and a MoS$_2$-WS$_2$ heterobilayer in (a)-(c) (top panels) and (d)-(f) (bottom panels), respectively. DFT is marked with black and SW+KC with red.}
\end{figure}

For lattice-matched systems, i.e. homobilayers and heterobilayers having identical chalgogen sites in both layers, which are developed by use of the HSSCs, it is of high importance that the resulting structures can accurately reproduce the electronic properties. In Fig. \ref{fig:lammps_fit}, a comparison between purely DFT calculated parameters and SW+KC can be seen.
Note, that $E_g$ shown in Fig. \ref{fig:lammps_fit}(a,d) is the energetically lowest momentum-conserving transition between the highest-lying valence band and the lowest-lying conduction band, which occurs at the $\vec{K}_{\pm}$ points for all stacking configurations and materials considered here.
LAMMPS does not provide $E_g$, instead this is calculated using DFT with the relaxed structures generated by our SW+KC force-field method. Note, that for the purpose of consistency, we adopt the notation that MoS$_2$-WS$_2$ implies that WS$_2$ lies above MoS$_2$ with respect to $z$.

In the case of homobilayers, the maximum deviation of $E_g$ is 22 meV, and occurs in the $\mathrm{H_M^M}$-stacking configuration. A similar maximum deviation of 25 meV is seen for the heterobilayer, which occurs in the $\mathrm{H_M^M}$-stacking configuration as well. For the remaining lattice-matched structures, the deviations are of similar magnitude. Fig. \ref{fig:lammps_fit} also demonstrates the high sensitivity of the bandgap with respect to changes in the structural degrees of freedom.

\begin{figure*}[ht]
\centering
\includegraphics[width=0.34\linewidth]{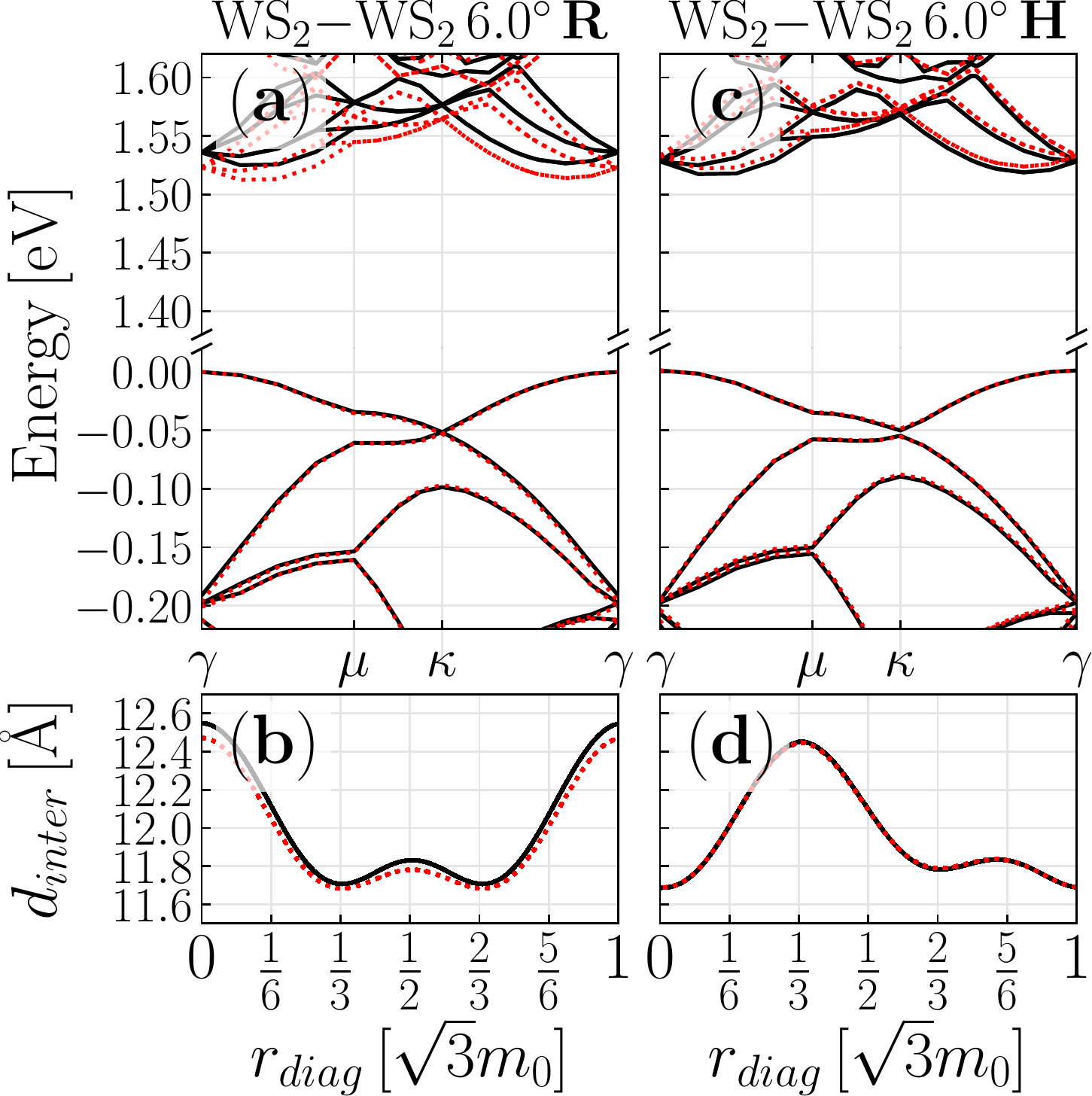}
\includegraphics[width=0.3168\linewidth]{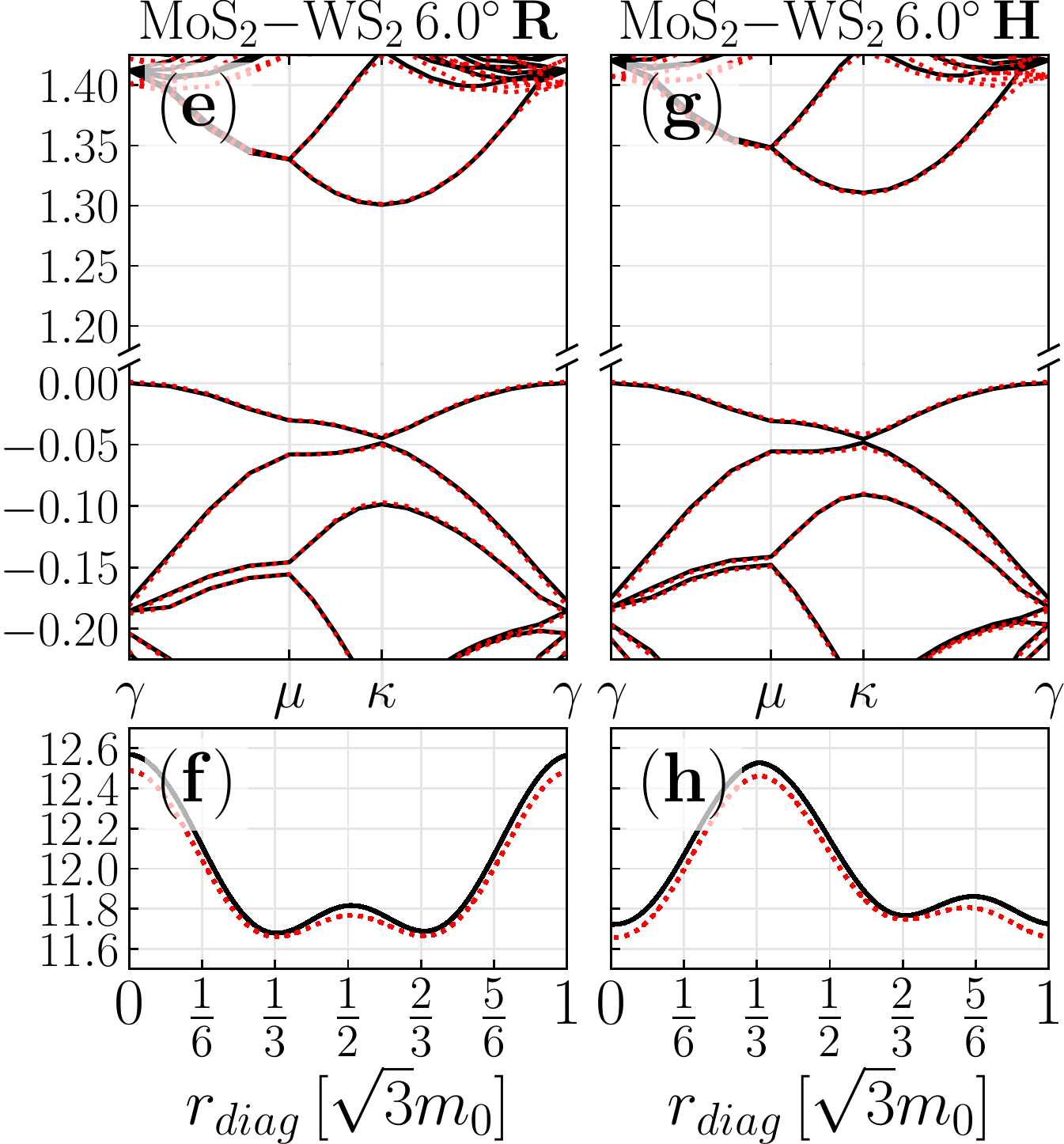}
\includegraphics[width=0.318\linewidth]{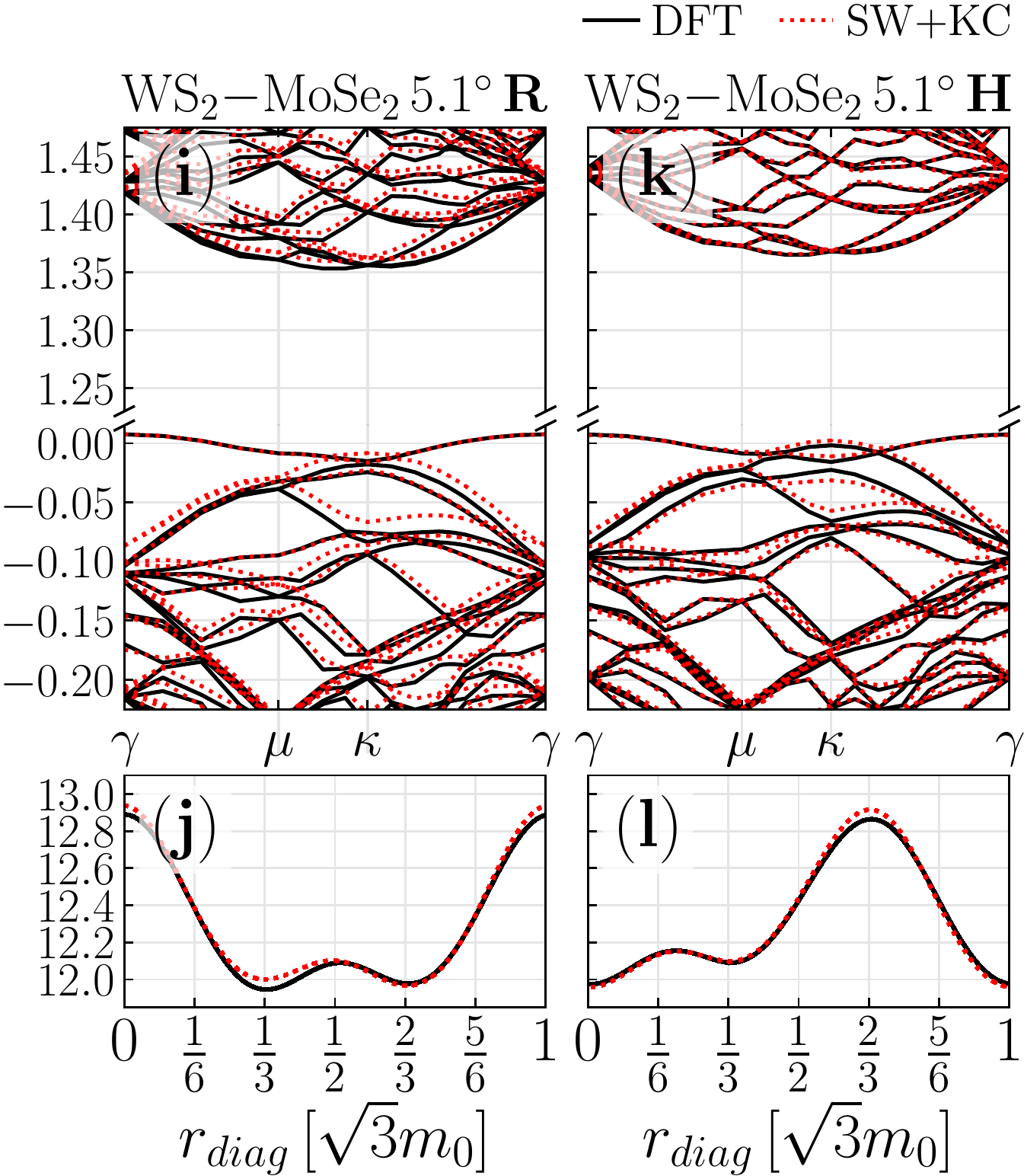}
\caption{Comparison of bandstructures and interlayer spacing profiles between DFT in solid black and SW+KC in dotted red. WS$_2$-WS$_2$ homobilayer with $\theta=6.0^\circ$ ($n_{atom} = 546$) in (a),(b) and (c),(d) for R- and H-stacking, respectively. MoS$_2$-WS$_2$ heterobilayer with $\theta=6.0^\circ$ ($n_{atom} = 546$) in (e),(f) and (g),(h) for R- and H-stacking, respectively. WS$_2$-MoSe$_2$ lattice-mismatched heterobilayer with $\theta=5.1^\circ$ ($n_{atom} = 642$) in (i),(j) and (k),(l) for R- and H-stacking, respectively. The bandstructures have the valence band maximum shifted to 0 in all cases, and the Greek indices ($\gamma$, $\mu$ and $\kappa$) denote the high-symmetry points of the moiré (mini) BZ (usually denoted $\Gamma$, M and K in the BZ of the monolayer/untwisted bilayer). The interlayer spacing is interpolated and plotted along the long diagonal of the unit cell.}
\label{fig:bands}
\end{figure*}

Having established the SW+KC parameters of lattice-matched systems using the HSSCs, we now tackle some larger moiré structures. As such, we use some medium-scale moiré structures as benchmarks. Fig. \ref{fig:bands} shows comparisons between DFT- and SW+KC-relaxed  band structures and interlayer spacing profiles for different material cases. Greek indices denote the corners of the mini Brillouin zone (BZ) associated with a moiré structured bilayer. The interlayer spacing is plotted along the long diagonal of the unit cell (see Fig. \ref{fig:moire_whole}), which has a length of $\sqrt{3}m_0$, where $m_0$ denotes the moiré cell lattice constant. Fig. \ref{fig:bands}(a)-(d) displays the case of a WS$_2$-WS$_2$ homobilayer twisted at 6.0$^{\circ}$. For R-stacking ((a),(b)), the bands are well represented using our SW+KC relaxed structure with only a 13 meV decrease of the bandgap, likely due to the slight interlayer spacing profile discrepancy. In the case of H-stacking ((c)-(d)), the same applies except the bandgap is a mere 5 meV larger compared to the DFT relaxed structure results. In Fig. \ref{fig:bands}(e)-(h), the case of a MoS$_2$-WS$_2$ heterobilayer with a twist angle of 6.0$^{\circ}$ is shown. For both R- and H-stacking ((e),(f) and (g),(h) respectively), an excellent agreement is obtained between DFT- and SW+KC-relaxed structures in terms of band character. For the higher lying conduction bands around the $\vec{\gamma}$ point, there is only a 10 meV discrepancy. We again attribute this to the slightly decreased interlayer spacing profiles of SW+KC in both cases, as seen in Fig. \ref{fig:bands}(f),(h).

For lattice-mismatched systems, the optimization of the KC parameters was performed for all possible combinations of metal and chalcogen atoms, as explained in the methodology section (see Table \ref{tab:mismatch}). A good agreement is obtained between the DFT- and SW+KC-relaxed structures for all lattice-mismatched cases. For the sake of brevity, only the case of a WS$_2$-MoSe$_2$ heterobilayer rotated at 5.1$^{\circ}$ is shown in Fig. \ref{fig:bands}(i)-(l). For R-stacking ((i),(j)), the highest lying valence band is only 7 meV higher than the DFT value at the $\vec{\kappa}$-point. The lowest-lying conduction band is only 6 meV above the DFT one. In general, we see small discrepancies between the valence and conduction bands for the DFT and SW+KC-relaxed structures below 20 meV. For H-stacking ((k),(l)), the valence bands are well described except for a 4 meV discrepancy of the highest-lying valence band near the $\vec{\kappa}$ point.

In general, we note that the slight difference in bandgap and band curvature between DFT and our SW+KC-relaxed moiré structures arise from small inaccuracies in the interlayer spacing profiles. Note, that this is not always the case with the KC-parameters presented by Ref. \cite{Naik2019}, where the binding energy was the target property. We also find that the accuracy of our lattice-matched SW+KC parameters reduce with growing twist angle. This is expected, since we fit to the untwisted HSSCs, which are not well represented in moiré structures with such low periodicity. Conversely, the parameters are expected to have better accuracy with decreasing twist angle. For angles below $~3^\circ$, where large-scale atomic reconstruction starts to appear, the accuracy of  methodology is still maintained and most properties are well captured, including the atomic reconstructions, as discussed in Sec. \ref{sec:reconstruction}.

\section{Approximating Moiré Potentials}

A defining feature of two-dimensional lattice-matched moiré structures is the spatial variation of local stacking order across the structure, leading to variation of local properties. Many combinations of TMDs possess type-II band alignment \cite{Ozcelik2016,Zhang2016}, and as such, the variation of the local bandgap at $\vec{K}_{\pm}$ across the structure will, for many purposes, describe the \textit{interlayer moiré potential} \cite{Geng2020,Wu2018,Gotting2022}. However, it is worth mentioning, that in the case of a large lattice-mismatch between the constituting layers, developing such a potential becomes non-trivial.

We propose two interpolation-based methods for calculating the interlayer moiré potential of lattice-matched systems. Moreover, any electronic property that can be identified locally, can be accessed in the moiré structure directly with these two methods, e.g. variation of the VBM, CBM etc.
In both methods, the moiré supercell is subdivided into small units the size of the monolayer unit cell, for which local properties can be calculated. The first method, which we call the high-symmetry interpolation method (HSIM), is based on the \textit{local high-symmetry stacking character} - a geometrical quantity that measures the similarity between the local stacking configuration within the moiré cell and the HSSCs. Being based only on the six HSSCs, computing the DFT-properties is fast and allows for high-throughput computations. It also allows for easy visualization of reconstructed domains. The second method, which we call the grid based interpolation method (GBIM), relies on computing the local properties using DFT not only for the HSSCs, but also every local stacking configuration in between, which can then be interpolated over the moiré supercell. In principle, this scheme is more precise, since it relies less on interpolation and more on \textit{ab initio} calculations. However, it is time consuming, as many DFT computations using different in-plane displacements and interlayer spacing are needed. In what follows, both methods are explained in detail and case studies are shown.

\subsection{\label{sec:HSIM}High-symmetry interpolation method (HSIM)}

\begin{figure}[ht]
\centering
\includegraphics[width=0.99\linewidth]{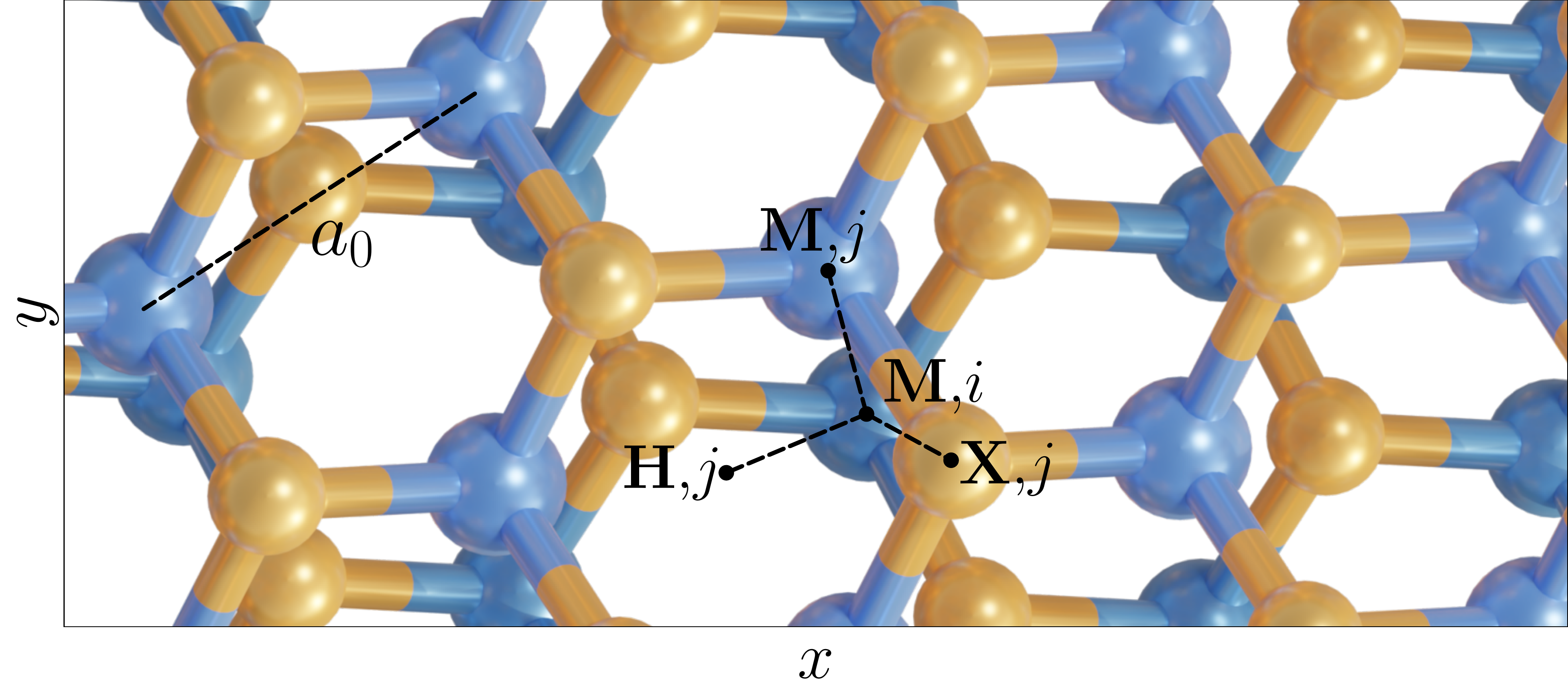}
\caption{\label{fig:method} Close-up of an R-stacked lattice-matched moiré structure for $\theta=6.0^{\circ}$. For a metal site $i$ in one layer, the nearest transverse metal site, $\mathrm{M},j$, chalcogen site, $\mathrm{X},j$, and hexagonal center, $\mathrm{H},j$, in the adjacent layer is seen.}
\end{figure}

For every metal site in one layer, $\vec{\rho}_{\mathrm{M},i} = (x_{\mathrm{M},i},y_{\mathrm{M},i})$, we find the transverse distance to the closest metal site in the adjacent layer, e.g. $d_{\mathrm{M},i}^{\mathrm{M}} = \mathrm{min}(|\vec{\rho}_{\mathrm{M},i} - \vec{\rho}_{\mathrm{M},j}|)$, where $j$ runs through every metal site in the adjacent layer (see Fig. \ref{fig:method}). The largest distance possible is $a_0 / \sqrt{3}$.

\begin{figure}[ht]
\centering
\includegraphics[width=0.99\linewidth]{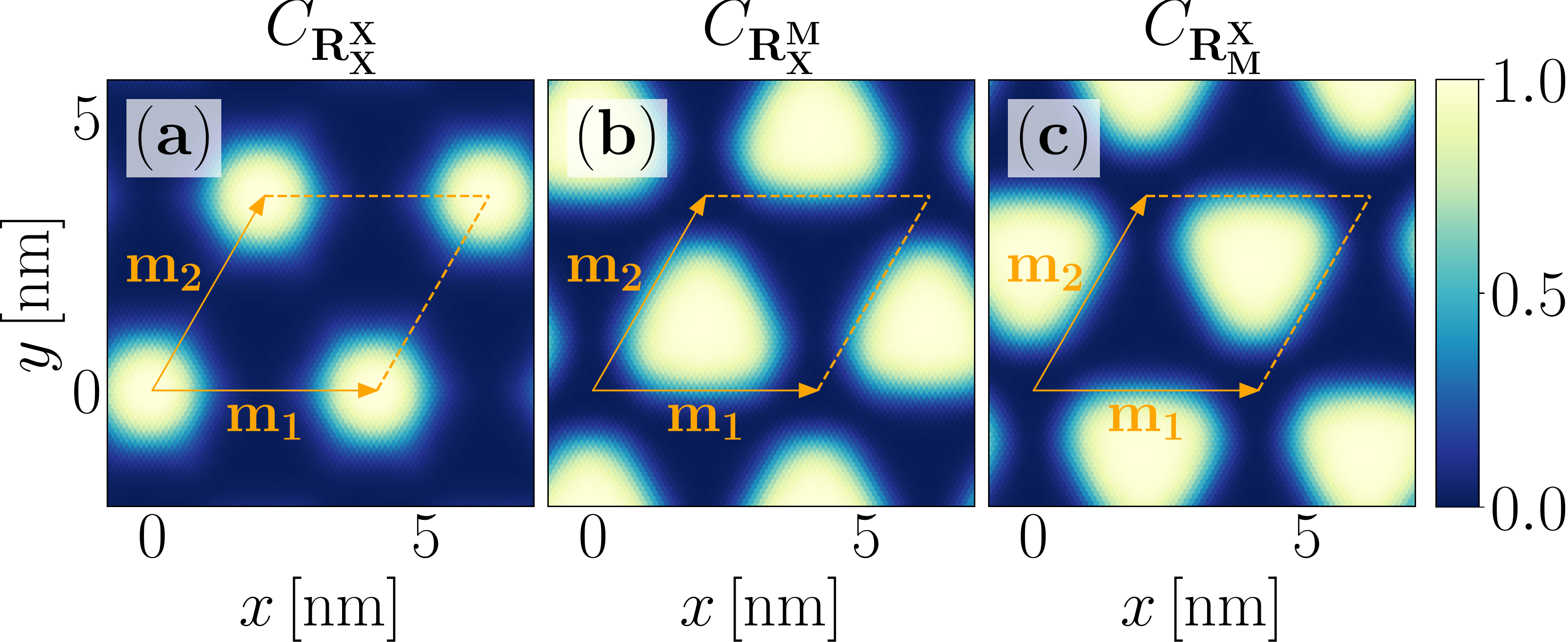}
\caption{\label{fig:coefficients}Variation of $C_{\mathrm{R_X^X}}$, $C_{\mathrm{R_X^M}}$ and $C_{\mathrm{R_M^X}}$ across an R-stacked MoS$_2$-WS$_2$ bilayer with $\theta=4.41^{\circ}$ in (a), (b) and (c), respectively. The remaining coefficients are 0. The structure was relaxed using LAMMPS.}
\end{figure}

As such, we can define the parameter $c_{\mathrm{M},i}^{\mathrm{M}}=1 - \sqrt{3}d_{\mathrm{M},i}^{\mathrm{M}} / a_0$, which is unity for perfectly aligned metal atoms, e.g. $\mathrm{R_X^X}$- and $\mathrm{H_M^M}$-stacking, and zero for the remaining HSSCs. Eight analogous parameters can be developed, e.g.
\begin{equation*}
    \{c_{\mathrm{S}_1,i}^{\mathrm{S}_2}(\vec{\rho}_{\mathrm{S}_1,i}) \quad \mathrm{for} \quad \mathrm{S}_1,\mathrm{S}_2 \in \{\mathrm{M},\mathrm{X},\mathrm{H}\}\},
\end{equation*}
where $\mathrm{X}$ and $\mathrm{H}$ denote chalcogen sites and hexagonal centers, respectively. For the purpose of consistency, it is assumed that $\mathrm{S}_2$ lies above $\mathrm{S}_1$ with respect to $z$. $\{c_{\mathrm{S}_1,i}^{\mathrm{S}_2}\}$ is then interpolated on a skewed grid that spans the moiré unit cell. Stacking coefficients are now found as
\begin{align*}
    C_{\mathrm{R_X^X}} = c_{\mathrm{M}}^{\mathrm{M}}c_{\mathrm{X}}^{\mathrm{X}}c_{\mathrm{H}}^{\mathrm{H}}&,\quad 
    C_{\mathrm{H_X^M}} = c_{\mathrm{X}}^{\mathrm{M}}c_{\mathrm{M}}^{\mathrm{X}}c_{\mathrm{H}}^{\mathrm{H}}, \\
    C_{\mathrm{R_X^M}} = c_{\mathrm{X}}^{\mathrm{M}}c_{\mathrm{H}}^{\mathrm{X}}c_{\mathrm{M}}^{\mathrm{H}}&,\quad 
    C_{\mathrm{H_X^X}} = c_{\mathrm{H}}^{\mathrm{M}}c_{\mathrm{X}}^{\mathrm{X}}c_{\mathrm{M}}^{\mathrm{H}}, \\
    C_{\mathrm{R_M^X}} = c_{\mathrm{H}}^{\mathrm{M}}c_{\mathrm{M}}^{\mathrm{X}}c_{\mathrm{X}}^{\mathrm{H}}&,\quad 
    C_{\mathrm{H_M^M}} = c_{\mathrm{M}}^{\mathrm{M}}c_{\mathrm{H}}^{\mathrm{X}}c_{\mathrm{X}}^{\mathrm{H}}.
\end{align*}
Finally, the stacking coefficients are normalized such that $\sum_nC_n({\vec{\rho}})=1$, where $n$ spans the HSSCs. $C_n$ is seen in Fig. \ref{fig:coefficients} for R-stacking. The $C_n$ with $n\in\{\mathrm{H_X^M},\mathrm{H_X^X},\mathrm{H_M^M}\}$ are all 0 in this case.
The next step is finding the interlayer spacing profile, $d_{inter}(\vec{\rho})$, where $\vec{\rho}=(x,y)$. Using the variation of $E_g$, $E_g(\vec{\rho})$, as an example, it can be seen that
\begin{equation}
    E_g(\vec{\rho}) = \sum_nC_n(\vec{\rho})E_g(n,d_{inter}(\vec{\rho})),
    \label{eq:HSIM}
\end{equation}
assuming the variation of $E_g$ with $d_{inter}$ is known for all HSSCs. Assuming that every subcell of the moiré structure can be described by a superposition of HSSCs is an approximation, but has the benefit of easy visualization of domains, as seen in Fig. \ref{fig:coefficients}. It shows great accuracy and $E_g(n,d_{inter})$ can be extracted within few calculations, making it quite fast to implement for all lattice-matched systems.

\subsection{Grid based interpolation method (GBIM)}

A more general implementation can be developed by using the untwisted bilayer with a transverse shift $\vec{\rho}_s=(x_s,y_s)$ between the layers, where $\vec{\rho}_s=0$ corresponds to either $\mathrm{R_X^X}$- or $\mathrm{H_M^M}$-stacking. We calculate $E_g^{\textrm{BL}}(\vec{\rho}_s,d_{inter})$, where $\vec{\rho}_s$ is the transverse distance between metal sites in each layer. Then, for a given lattice-matched moiré system, for metal site $i$ in one layer, we can find the vector $\vec{\rho}_i = \vec{\rho}_{\mathrm{M},j} - \vec{\rho}_{\mathrm{M},i}$, where $j$ denotes the index of the nearest metal site in the adjacent layer. Then, the value of $E_g$ at metal site $i$ is simply
\begin{equation}
    E_g(\vec{\rho}_{\mathrm{M},i}) = E_g^{\textrm{BL}}(\vec{\rho}_i,d_{inter}(\vec{\rho}_{\mathrm{M},i})).
\end{equation}
Note, that $\vec{\rho}_i$ should be adjusted relative to the rotation of the individual layers, since the layers will likely be slightly angled compared to the systems used in computing $E_g^{\textrm{BL}}(\vec{\rho}_s,d_{inter})$. Finally, $E_g$ is interpolated over the entire moiré unit cell.

\begin{figure}[ht]
\centering
\includegraphics[width=0.99\linewidth]{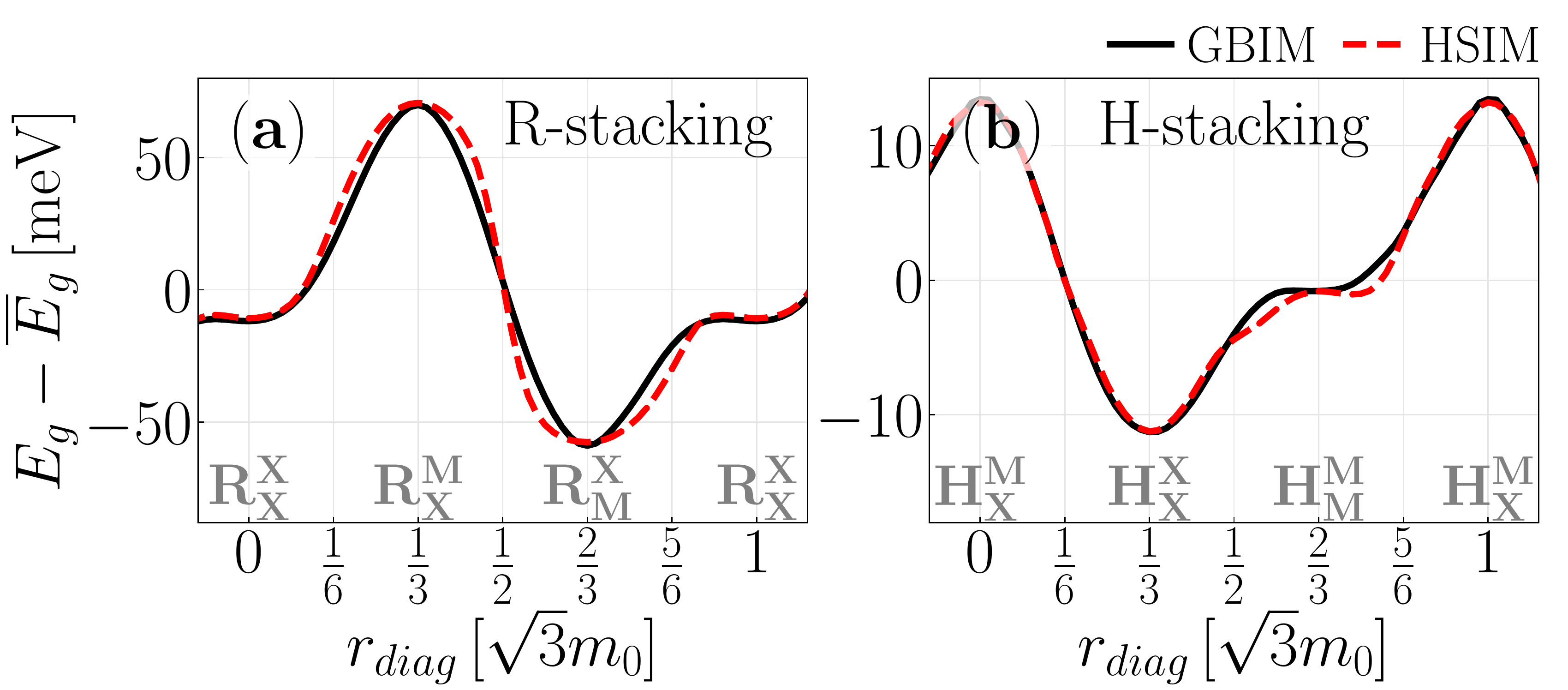}
\caption{\label{fig:hsymvsgrid}Variation of $E_g$ along the long diagonal of an R- and H-stacked MoS$_2$-WS$_2$ bilayer with $\theta=4.41^{\circ}$ in (a) and (b), respectively. The solid black and dashed red curves represent bandgap variation found using the GBIM and HSIM, respectively.}
\end{figure}

In principle, the GBIM should be more accurate than the HSIM, but is also computationally more expensive. We use twelve steps for $x_s$ and $y_s$ combined with sixteen increments for $d_{inter}$ when tabulating $E_g^{\textrm{BL}}(\vec{\rho}_s,d_{inter})$. This translates to 4608 separate DFT calculations to cover R- and H-stacking for one material, whereas the HSIM needs only 96. In Fig. \ref{fig:hsymvsgrid}, a comparison between the HSIM and the GBIM can be seen for the variation of $E_g$ in MoS$_2$-WS$_2$ with $\theta=4.41^{\circ}$. At the high-symmetry points, both methods yield the same value as expected, but the HSIM is slightly inaccurate in between.

\section{\label{sec:reconstruction}Atomic Reconstruction and\protect\\ Energetic Landscape}

As mentioned, the energetic landscape of 2D moiré structures is constituted by three codependent factors: the local stacking arrangement, the associated interlayer spacing, and the atomic reconstruction. Often, the latter two, being relaxation effects, are not considered in simulations \cite{Wu2017,Wu2018,Gotting2022,Rademaker2022,Yu2017,Hichri2021,Lagoin2021,Tran2019,LiLu2021,Zheng2021,Pan2020_2,Zhai2020,Brem2020,Wu2019,WuTutuc2018}, but can be managed with SW+KC force-field relaxation.

\begin{figure}[ht]
\centering
\includegraphics[width=0.99\linewidth]{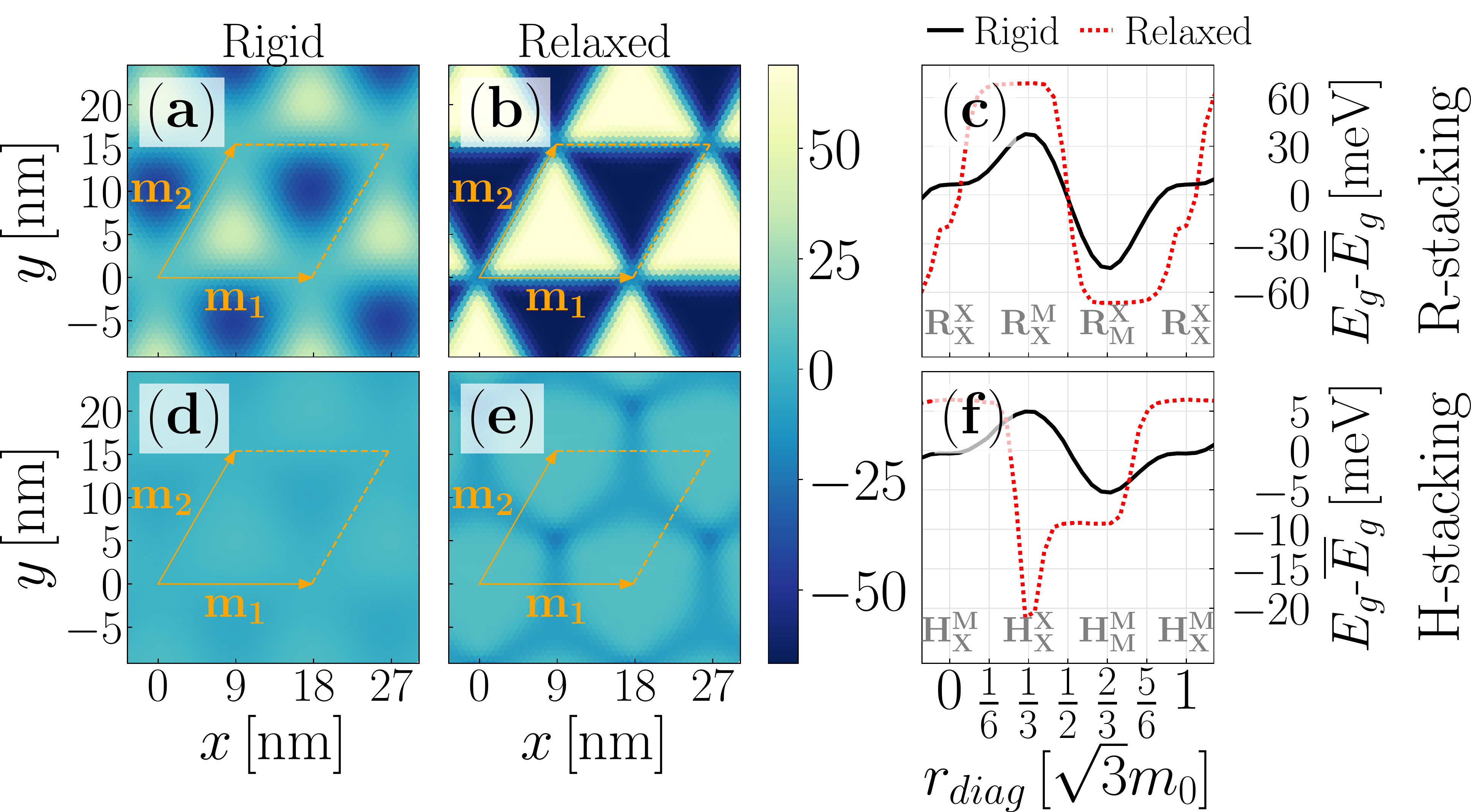}
\caption{\label{fig:moire_pot}Variation of $E_g$ at $\vec{K}_{\pm}$ across a 1.01$^{\circ}$ twisted R-stacked MoS$_2$-WS$_2$ bilayer without and with relaxation effects in the (a),(d) and (b),(e), respectively. Comparison between the two cases along the long diagonal of the moiré unit cells in (c),(f). (a)-(c) and (d)-(e) represent R- and H-stackings, respectively.}
\end{figure}

For MoS$_2$-WS$_2$, which possesses type-II band alignment \cite{Ozcelik2016,Zhang2016}, the interlayer moiré potential is often described as the spatial variation of the local bandgap at $\vec{K}_{\pm}$. In Fig. \ref{fig:moire_pot}, the variation of $E_g - 
\overline{E}_g$ across an R- and H-stacked MoS$_2$-WS$_2$ bilayer with $\theta=1.01^{\circ}$ is seen, where $\overline{E}_g$ is the mean value across the unit cell. In the rigidly twisted case, the average interlayer spacing of the three R- or H-stacked HSSCs are used as interlayer spacing for Fig. \ref{fig:moire_pot}(a) and Fig. \ref{fig:moire_pot}(d), respectively.
The discrepancy between modeling the potential with- and without relaxation effects is apparent. In the case of R-stacking, which has larger potential depth than H-stacking, the depth with- and without relaxation effects are here estimated to be 80 meV and 135 meV, respectively. For H-stacking, these numbers are much lower, namely 10 meV and 28 meV for rigid and relaxed, respectively. As a consequence, phenomena such as exciton trapping may be realized more easily in R-stacked systems than H-stacked analogs. Interestingly, for H-stacking, the minimum of the potential resides in the $\mathrm{H^X_X}$-domain post-relaxation as opposed to the $\mathrm{H^M_M}$-domain pre-relaxation. Finally, the effect of atomic reconstruction also greatly changes the relative widths of the potential wells, resulting in a sharper and more well-defined potential. We conclude that atomic reconstruction significantly alters the range of $\theta$ in which exciton trapping occurs.

In Table \ref{tab:binding}, the \textit{interlayer binding energy per atom}, $E_b$, found as $E_b=(E_{\mathrm{MoS_2}\text{-}\mathrm{WS_2}} - E_{\mathrm{MoS_2}} - E_{\mathrm{WS_2}})/6$ is shown, where $E_{\mathrm{MoS_2}\text{-}\mathrm{WS_2}}$ is the total energy of the untwisted bilayer system, and $E_{\mathrm{MoS_2}}$ and $E_{\mathrm{WS_2}}$ denote the total energies of constituting monolayers found separately. As mentioned, the discrepancy in $E_b$ between DFT and SW+KC is expected, since this was not the target property during development of our KC parameters.
\begin{table}[ht]
\caption{\label{tab:binding} Binding energy of MoS$_2$-WS$_2$ in the six high-symmetry stacking configurations from DFT and from SW+KC.}
\begin{ruledtabular}
\begin{tabular}{ccccccc}
$E_b$ (meV)     & $\mathrm{R^X_X}$  & $\mathrm{R^X_M}$  & $\mathrm{R^M_X}$  & $\mathrm{H^M_X}$  & $\mathrm{H^X_X}$  & $\mathrm{H^M_M}$  \\ 
\hline
DFT             & -21.8             & -34.3             & -34.5             & -34.6             & -22.4             & -31.6             \\
\hline
SW+KC           & -25.4             & -44.4             & -44.6             & -44.7             & -28.9             & -38.4             \\
\end{tabular}
\end{ruledtabular}
\end{table}

For R-stacking, the nearly identical $E_b$ of the $\mathrm{R^X_M}$- and $\mathrm{R^M_X}$-configurations facilitates a simultaneous growth of these domains with decreasing $\theta$ (i.e. large moiré unit cells), while the opposite is true for $\mathrm{R^X_X}$, explaining the formation of a mesh of triangular domains, as seen in Fig. \ref{fig:moire_pot}(b). For H-stacking, the $\mathrm{H^M_X}$-configuration is energetically favorable, resulting in hexagonal domains with decreasing $\theta$. The $\mathrm{H^M_M}$-like domains shrink slower than those associated with $\mathrm{H^X_X}$, as seen from the associated $E_b$ (see also \cite{Halbertal2021}).

Lastly, $E_b(\theta)$ can be considered in order to access the formation rate of domains. Using the HSIM, the variation of the local $E_b$ across a moiré unit cell can be approximated, and the mean can be used to approximate $E_b$ of the moiré unit cell, albeit neglecting the effects of strain imposed by atomic reconstruction and corrugation from the varying interlayer spacing. In the case of pure DFT, $E_b$, is found directly as
\begin{equation}
    E_b = E_{\text{moiré}} - (E_{\mathrm{MoS_2}}+E_{\mathrm{WS_2}})/2,
    \label{eq:binding}
\end{equation}
where all energies are divided by the number of atoms, and $E_{\text{moiré}}$ denotes the total energy per atom of the moiré structure. However, $E_b$ has contributions from the strain imposed by layer corrugation and atomic reconstruction. The energy associated with these effects is denoted $E_{\text{corr}}$ and is not captured by the HSIM. Instead, the $E_b$ found by the HSIM should be compared to
\begin{equation}
    E_b - E_{\text{corr}} = E_{\text{moiré}} - (E_{\mathrm{MoS_2,\text{moiré}}}+E_{\mathrm{WS_2,\text{moiré}}})/2,
\end{equation}
where $E_{\mathrm{MoS_2,\text{moiré}}}$ and $E_{\mathrm{WS_2,\text{moiré}}}$ denote the total energy per atom for the corrugated and reconstructed constituting monolayers. This is computed in separate DFT calculations having half the number of atoms as the moiré structure they constitute.

\begin{figure}[ht]
\centering
\includegraphics[width=0.99\linewidth]{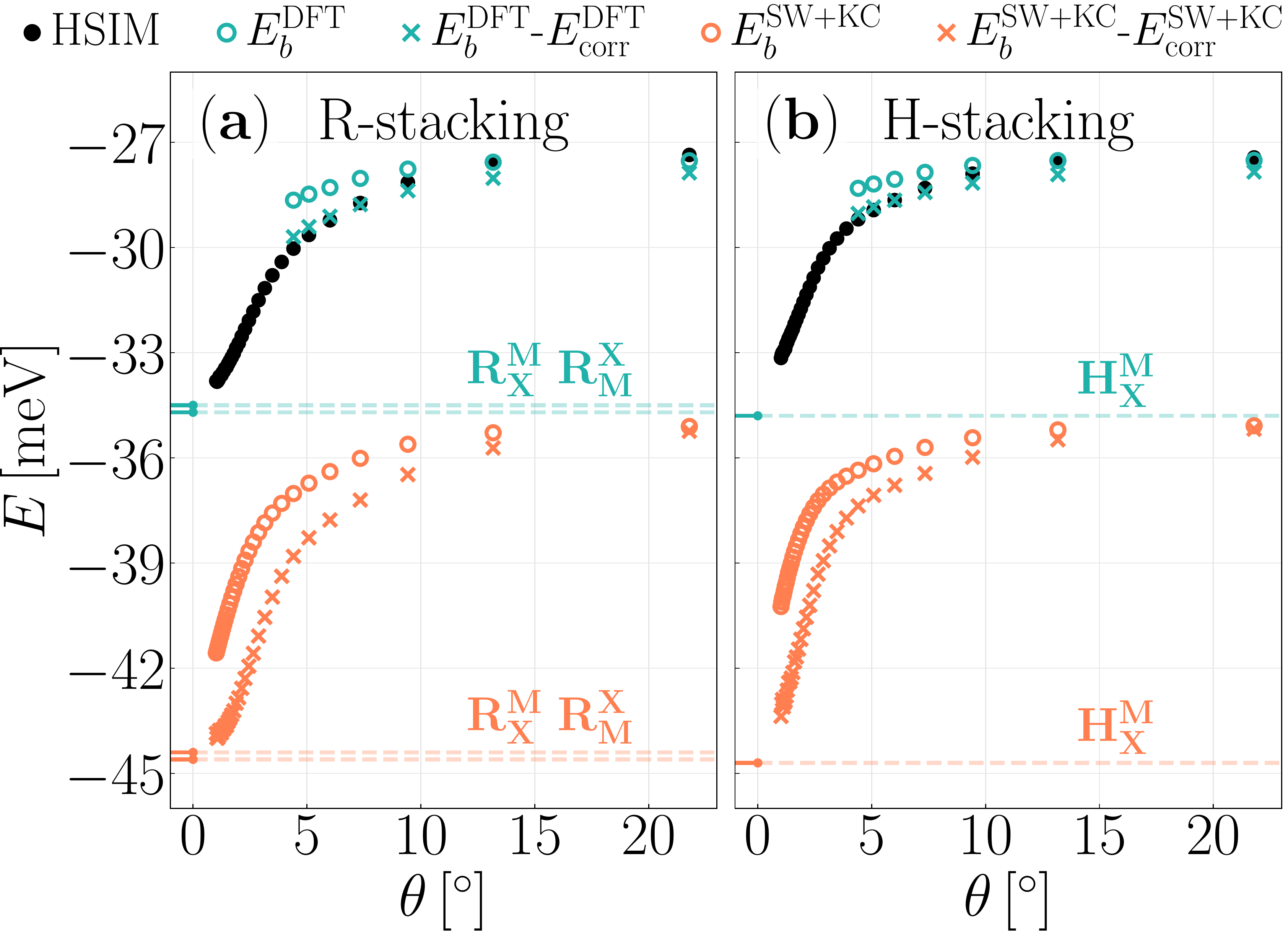}
\caption{\label{fig:binding}$E_b(\theta)$ for MoS$_2$-WS$_2$. (a) and (b) are for R- and H-stacking, respectively. Black points are found using the HSIM on the SW+KC-relaxed structures, but with DFT-based parametrization of the HSIM as seen in Eq. \eqref{eq:HSIM}.}
\end{figure}

With SW+KC, $E_b$ is found analogously to Eq. \eqref{eq:binding}, but $E_{\text{corr}}$ is found directly by comparing the energy of the SW-potential in the two layers to that of the constituting rigid monolayers. The variation of these quantities with $\theta$ is seen in Fig. \ref{fig:binding}.

A common feature for all energy scales in Fig.~\ref{fig:binding} is the tendency towards the value of the stable configurations for $\theta\rightarrow0$. For vanishing $\theta$, the relative size of the domain walls becomes negligible. As such, $E_{\text{corr}}$ should vanish in the limit of vanishing $\theta$. The faster convergence towards the $E_b$ of $\mathrm{R_X^M}$/$\mathrm{R_M^X}$ for R-stacking indicates that the triangular domains form more rapidly with decreasing $\theta$ compared to the hexagonal $\mathrm{H_X^M}$-domains of H-stacking. Although the values of $E_b^{\mathrm{SW+KC}}$ and $E_{\text{corr}}^{\mathrm{SW+KC}}$ may appear off scale, they illustrate the tendencies faithfully. Additionally, the graph of $E_b^{\mathrm{DFT}}-E_{\text{corr}}^{\mathrm{DFT}}$ serves as a benchmark, showing that the HSIM has accuracy within the 0.5 meV range, and further that $E_b$ of SW+KC relaxed structures can be recovered to agree with DFT.

\begin{figure}[ht]
\centering
\includegraphics[width=0.99\linewidth]{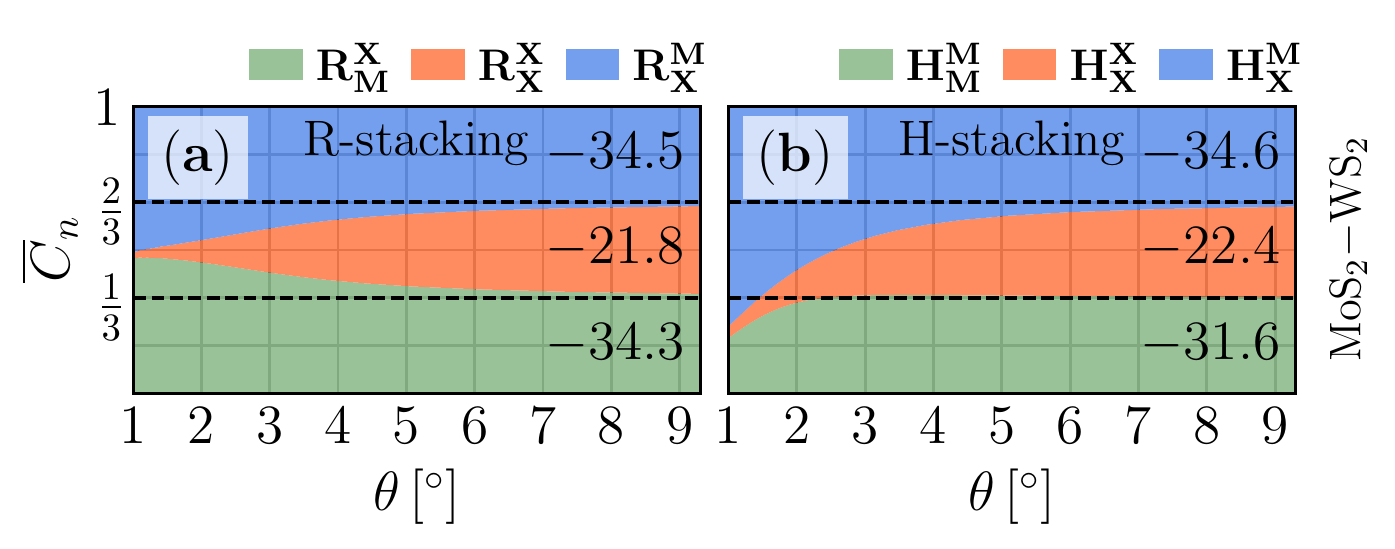}
\caption{\label{fig:reconstruction1}The mean of $C_n$ over the unit cell with respect to the twist angle, $\theta$, of MoS$_2$-WS$_2$ found using SW+KC and the HSIM. (a) correspond to R-stacking such that only $\mathrm{R_X^X}$, $\mathrm{R_M^X}$ and $\mathrm{R_X^M}$ are non-zero and vice versa for (b). For each stacking configuration, the corresponding $E_b$ of untwisted MoS$_2$-WS$_2$ is shown in meV.}
\end{figure}

%\gbnote{check following paragraph carefully, I have rewritten}
Fig. \ref{fig:reconstruction1} shows the mean of the stacking coefficients $C_n$ over the moiré unit cell of MoS$_2$-WS$_2$, which can be computed using the HSIM as described in Sec.~\ref{sec:HSIM}. $C_n$ represents the normalized contributions of the different stacking configurations to the fully relaxed (reconstructed) moiré structure.
% Large angle
For R-stacking (H-stacking), the three possible domains are: $\mathrm{R_M^X}$ (green), $\mathrm{R_X^X}$ (red), $\mathrm{R_X^M}$ (blue) ($\mathrm{H_M^M}$ (green), $\mathrm{H_X^X}$ (red), $\mathrm{H_X^M}$ (blue)).
For larger angles, the fraction of the unit cell area occupied by each of the three domains is about $1/3$ for both R- and H-stacking.
% R-stacking: two domains, alrady at 1 Deg.
At an angle of 1$^\circ$ the structure for R-stacking (Fig.~\ref{fig:reconstruction1}(a)) is already reconstructed in such a way that the energetically less favorable $\mathrm{R_X^X}$ (red) domains represent only 2,5\% of the structure. Both $\mathrm{R_M^X}$ (green) and $\mathrm{R_X^M}$ (blue) domains are energetically equivalent, and hence, occupy roughly 50\% of the structure in the limit of small $\theta$. 
% H-stacking: one domain, but at smaller angles
For H-stacking (Fig.~\ref{fig:reconstruction1}(b)), at the same angle of 1$^\circ$, the less favorable $\mathrm{H_X^X}$ and $\mathrm{H_M^M}$ domains have significantly reduced contributions compared to the favorable $\mathrm{H_X^M}$ region, but $\mathrm{H_M^M}$ still represents ~20\% of overall structure.

% Conclusion
Fig.~\ref{fig:reconstruction1} allows us to draw quantitative conclusions on the angle dependence of the reconstruction effect. Indeed, neglecting reconstructions would lead to a constant equal proportion of all three coexisting stackings (dotted lines in Fig.~\ref{fig:reconstruction1}). In the case of R-stacking the reconstruction is nearly complete at an angle of 1$^\circ$, i.e., the moire structure is made of basically two type of low energy domains ($\mathrm{R_M^X}$ (green), $\mathrm{R_X^M}$ (blue)) separated by a very narrow $\mathrm{R_X^X}$(red) energetically unfavorable domain. For H-stacking at 1$^\circ$, the less favorable $\mathrm{H_M^M}$ (green) domain still covers 15-20\% of the area. 

\begin{figure}[ht]
\centering
\includegraphics[width=0.99\linewidth]{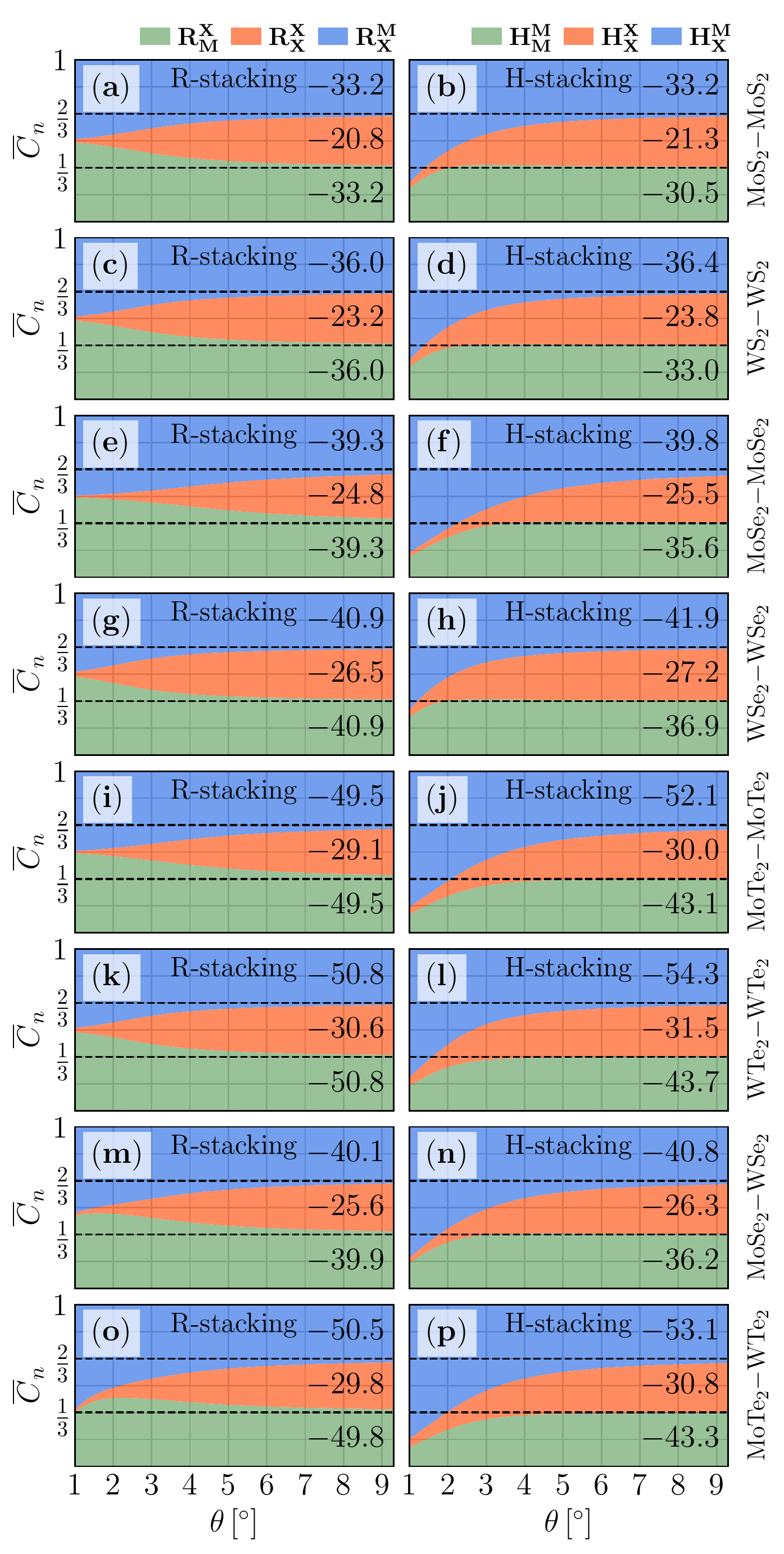}
\caption{\label{fig:reconstruction2}The mean of $C_n$ over the unit cell with respect to the twist angle, $\theta$, of eight lattice-matched bilayers found using SW+KC and the HSIM. Left panels correspond to R-stacking such that only $\mathrm{R_X^X}$, $\mathrm{R_M^X}$ and $\mathrm{R_X^M}$ are non-zero and vice versa for right panels. For each material and stacking configuration, the corresponding $E_b$ of the untwisted bilayer is shown in meV.}
\end{figure}

Fig. \ref{fig:reconstruction2} shows the same graph as Fig. \ref{fig:reconstruction1} for the remaining eight lattice-matched materials. Generally, all R-stacked materials (left panels of Fig. \ref{fig:reconstruction2}) display a simultaneous growth of $\mathrm{R_M^X}$ and $\mathrm{R_X^M}$ with decreasing $\theta$ except for MoTe$_2$-WTe$_2$, which can be attributed to the discrepancy in $E_b$ for these stacking configurations. We conclude that for both stackings and all materials considered here, except for MoSe$_2$-MoSe$_2$, that atomic reconstruction becomes especially prominent below an angle of 4-5$^\circ$. For MoSe$_2$-MoSe$_2$, atomic reconstruction occurs for angles below 6-7$^\circ$.

\section{Conclusion}

In conclusion, we have shown the dramatic consequences of incorporating relaxation effects on the \textit{interlayer moiré potential} of MoS$2$-WS$_2$. For R-stacking, this becomes about twice as deep at about 135 meV, and, for small angles, much wider. For H-stacking, the potential depth is nearly tripled, however, the width of the potential minima is still narrow, since it corresponds to the energetically unfavorable $\mathrm{H^X_X}$-configuration. Moreover, we have quantified the formation rate of domains due to atomic reconstruction for nine lattice-matched TMD moiré systems, and conclude that, in general, atomic reconstruction becomes prominent for $\theta$ smaller than 4-5$^{\circ}$, but does so in a continuous manner.

Furthermore, we have presented a methodology for developing KC-parameters for lattice-matched and -mismatched systems, and have developed such parameters for TMD moiré heterostructures. The method shows excellent agreement between DFT-calculated structural parameters and SW+KC-relaxed ones, which is further reflected in the bandstructure and the \textit{interlayer binding energy} with twist angle dependence. The force-field parameters along with a variety of relaxed structures can be found via Ref. \cite{ff_params}. We have further shown two methods for capturing moiré induced fluctuations of local properties in lattice-matched systems that do not require extensive \textit{ab initio} treatment. These methods allow for visualization of the importance of relaxation effects and further serve as a first step in developing accurate moiré potentials. However, further investigation is required to develop analogous tools for lattice-mismatched moiré structures. 

%Although the variation of $E_g$ provides an intuitive glimpse of the \textit{interlayer moiré potential}, it would be interesting to study the excited states of such a system from an \textit{ab initio} standpoint assisted by force-field relaxation.

In summary, starting from the force-field model, it is now possible to tackle excited state physics incorporating relaxation effects i.e. layer corrugation and atomic reconstruction. For models such as tight-binding, this was not possible before, and for \textit{ab initio} studies, the cumbersome first step of relaxation can be skipped, thus saving computational resources and time. Furthermore, a thorough dissection of the formation rate of domains with decreasing angle is required to gain quantitative insight into the mechanisms behind it.

\begin{acknowledgments}
The project is supported by the Deutsche Forschungsgemeinschaft (DFG) within the Priority Program SPP2244 2DMP and by the Cluster of Excellence ``Advanced Imaging 
of Matter'' of the Deutsche Forschungsgemeinschaft (DFG) -- EXC 2056 -- project ID 390715994.
\end{acknowledgments}

\bibliography{ForceFieldMoire}% Produces the bibliography via BibTeX.

\end{document}